\documentstyle [12pt]{article}

\def\be{\begin{equation}}
\def\ee{\end{equation}}
\def\bea{\begin{eqnarray}}
\def\eea{\end{eqnarray}}

\begin{document}

\begin{titlepage}

\title{Theory of  fields in quantized spaces}
\author{A.N. Leznov\\
Universidad Autonoma del Estado de Morelos,\\ 
CIICAp,Cuernavaca, Mexico}

\maketitle

\begin{abstract}

It is shown how the theory of the fields can be constructed in a consistent way in quantized spaces. All constructions are connected with unitary irreducible representations of real forms
of six dimensional rotation algebras $O(1,5),O(2,4),O(3,3)$ or equivalent real forms of four-dimensional unitary algebras $SU^*(4),SU(2,2),SL(4)$. The most surprising  fact is that that all operations in the construction of such field theories are connected with compact manifolds. 
The gauge invariant theory in quantized space is proposed. 
\end{abstract}

\end{titlepage}


\section{Introduction}

The goal of the present paper  is to show that field theory in quantum space may constructed on the same level of consistency as in the usual four-dimensional space-time formalism. 
The problem of the investigation below may formulated in a few words as follows. There exists some algebra connecting $15$ objects; four co-ordinates $x_i$, four momenta $p_i$, six generators of Lorentz transformations $F_{ij}$ and a unit "element" $I$. It is possible to realize such algebra, i.e. construct its irreducible representations. After this all operations
-translations (momenta) $p_i$, Lorentz transformations $F_{ij}$, operators of co-ordinates $x_i$ and the unit element $I$-are well defined  mathematically. In the usual case as a result of the resolution of such  an algebra we obtain  the well known expressions of these objects in terms of the operations of multiplication and differentiation. After this the rules of construction of the physical theory are the following (of course very roughly); it is necessary to enumerate all possible equations and systems invariant with respect to the group of the motion generated by the infinitesimal operators $p_i,F_{ij}$. In the case of the quantum space it is possible to conserve absolutely all ingredients of the above scheme. Such a procedure for the construction of a field theory in quantum space was proposed in \cite{3}.

We will work below with the algebra of  a quantum space proposed in \cite{1}  containing 3 dimensionful parameters of the square of the length $L^2$, square of momentum $M^2$ and action
$H$. Commutation relations of this algebra take the following form
$$
[p_i,x_j]=ih(g_{ij}I+{F_{ij}\over H}),\quad  [p_i,p_j]={ih\over L^2}F_{ij},
\quad [x_i,x_j]={ih\over M^2}F_{ij},
$$
\be
[I,p_i]=ih({p_i\over H}-{x_i\over L^2}), \quad [I,x_i]=ih({p_i\over M^2}-
{x_i\over H}),\quad [I,F_{ij}]=0 \label{2}
\ee
$$
[F_{ij},x_s]=ih(g_{js}x_i-g_{is}x_j),\quad [F_{ij},p_s]=ih(g_{js}p_i-g_{is}p_j)
$$
$$
[F_{ij},F_{sk}]=ih(g_{js}F_{ik}-g_{is}F_{jk}-g_{jk}F_{is}+g_{ik}F_{js})
$$
and obviously have the correct limit under $L^2\to \infty,M^2\to \infty,H\to \infty$
to the usual theory. In \cite{1} it was shown that  the dependence between the dimensionful parameters of the algebra of the quantum space is equivalent to one of the algebras of six-dimensional groups of rotations $O(1,5),O(2,4),O(3,3)$ or some its degenerate forms.
\newpage
{\hspace*{12.5cm}\small Table 1. \quad{} }
{\small \parbox[t]{13.9cm}
{\hspace*{1.5cm}\small The real simple Lie algebras that correspond to 
the different values of  $L^2$,\\\hspace*{1.5cm} $M^2$ and  $H^2$ parameters.}}
\smallskip

\begin{center}
\begin{tabular}{|l|c|}
\hline
\qquad \quad $M^{2}$,\quad $L^{2}$ and\quad $H^{2}$  values & Algebra\quad 
\\[5pt]\hline\hline
\quad $H^{2}<M^{2}L^{2}$, \quad $M^{2}>0$, \quad $L^{2}>0$ & \qquad $%
o(2,4)\qquad $ \\[5pt]\hline
\quad $H^{2}<M^{2}L^{2}$, \quad $M^{2}<0$, \quad $L^{2}<0$ & \qquad $%
o(2,4)\qquad $ \\[5pt]\hline
\quad $M^{2}>0$,\quad $L^{2}<0$, \quad or\quad $M^{2}<0$,\quad $L^{2}>0$
\quad  & \qquad $o(2,4)\qquad $ \\[5pt]\hline
\quad $H^{2}>M^{2}L^{2}$, \quad $M^{2}>0$, \quad $L^{2}>0$ & \qquad $%
o(1,5)\qquad $ \\[5pt]\hline
\quad $H^{2}>M^{2}L^{2}$, \quad $M^{2}<0$, \quad $L^{2}<0$ & \qquad $%
o(3,3)\qquad $ \\[5pt]\hline
\end{tabular}
\end{center}

\smallskip

For $H^{2}=M^{2}L^{2}$ and $M^{2}>0$, $L^{2}>0$ the $o(1,5)$ algebra
degenerates into a semidirect product of the $o(1,4)$ algebra and the algebra
of 5 - translations, while for $H^{2}=M^{2}L^{2}$ 
and $M^{2}<0$, $L^{2}<0$ the $o(3,3)$ algebra degenerates into a semidirect 
product of the $o(2,3)$ algebra and the algebra of 5 - translations. 
Note that a transition to the limit $A_{\alpha }\rightarrow \infty $, 
where $A_{\alpha }$ is any term of the set $\{M^{2},$ $L^{2},$ $H^{2},$ $(M^{2},L^{2})\}$, 
do not exclude the algebras (\ref{finalg}) from the class of simple algebras, 
as distinct from the transitions $B_{\alpha }\rightarrow \infty $, where $B_{\alpha }\in $ 
$\{(M^{2},H^{2}),$ $(L^{2},H^{2}),$ $(M^{2},L^{2},H^{2})\}$ or $M^{2}L^{2}\rightarrow H^{2}$.

Their most important property is   that all irreducible representations of such real forms may realized on the compact manifolds corresponding to $O(5),O(2)\times O(4)$, or $ O(3)\times O(3)$. And thus all events in "real" (quantum space) world are in some sense only projections from the motion on such compact manifolds. Furthermore all operations of differentiation in the description  of differential equations are dependent upon on infinitesimal motions on compact manifolds. 

The explicit representation of the commutation relations (\ref{2}) it terms of dimensionless generators of the six-dimensional orthogonal algebra $F_{i,j}$ $(1\leq i,j\leq 6)$
$$
[T_{ij},T_{sk}]=\delta_{js}T_{ik}-\delta_{is}T_{jk}-\delta_{jk}T_{is}+\delta_{ik}T_{js})
$$
is the following one:
\be
I=i\nu T_{65},\quad x_i=AT_{6,i}+BT_{5,i},\quad p_i=CT_{6,i}+DT_{5,i},\quad F_{i,j}=ihT_{i,j}
\label{REZ}
\ee
where $\nu^2=-({h\over H})^2({H^2\over M^2L^2}-1),A={L^2\over H}C-{L^2\over h}\nu D, B={L^2\over H}D+{L^2\over h}\nu C$, and $C^2+D^2={h^2\over L^2}$. 

Inverting the last relations  leads to the following expressions of the generators
$T_{ij}$ in terms of the physical variables
$$
T_{6,5}=-{i\over \nu}I,\quad T_{i,j}=-{i\over h} F_{i,j}
$$
$$
T_{5,i}=-{i\over \nu h}D x_i+({iL^2\over \nu hH}D+{L^2\over h^2}C)p_i
$$
$$
T_{6,i}={i\over \nu h}C x_i-({iL^2\over \nu hH}C+{L^2\over h^2}D)p_i
$$

Three Casimir operators  are  represented in the following form
$$
K_2={\rm spur} (T^2),\quad K_3=\sum \epsilon_{ijklmn}T_{ij}T_{kl}T_{mn},\quad K_4={\rm spur} (T^4)
$$
or in other words
$$
K_2=\sum_{i\leq j} T_{i,j}T_{i,j},
$$
$$
K_3=T_{6,5}\sum_{1\leq i\leq j\leq k\leq l\leq 4} \epsilon_{ijkl} T_{ij}T_{kl}+\sum_{1\leq i\leq j\leq k\leq l\leq 4}\epsilon_{ijkl}T_{6,i}T_{5,j}T_{i,j}
$$

The most simple way for understanding the content of the table above and the explicit formulae connecting observable variables with generators of six-dimensional rotational algebra lies in 
the consideration of the explicit form of the second Casimir operator of the algebra (\ref{2}) in the terms of physical observables  
$$
K_2=-I^2+{1\over L^2}x_i^2+{1\over M^2}p^2_i-{p_ix_i+x_ip_i\over H}+
(f^2-l^2)(-{1\over H^2}+{1\over L^2M^2})=
$$
$$
\nu^2(-{I^2\over \nu^2}+{(x_i-{L^2\over H}p_i)^2\over \nu^2 L^2}+{L^2\over H^2}p_i^2+{(f^2-l^2)\over H^2})
$$
where  $a^2=a_1^2+a_2^2+a_3^2-a_4^2$,\  $l^2=(\vec l)^2,\ f^2=(\vec f)^2$ are squares of three dimensional generators of rotations and Lorentz boosts respectively. An operator of three-dimensional rotations is a compact generator and thus all generators (more exactly their squares) with the same sign in $K_2$ are always compact,but with the opposit sign are non-compact.In the case of $0 \leq L^2,M^2,\nu^2$ this is $O(3,3)$ algebra and so on. And the resolution of (\ref{REZ}) becomes absolutely clear. 

Two other   Casimir operators in three and four dimensions in the obsevable variables read as
$$
K_3=I(\vec f \vec l)-(\vec f [\vec p,\vec x])-(\vec l, (p_4\vec x-x_4\vec p))
$$
A non-trivial explicit expression for the Casimir operator of the fourth order may be obtained after non obvious modification of the corresponding operator of Casimir (of the fourth order) in the case of $O(5)$ algebra \cite{LMOD} 
$$
K_4=\sum_{i\leq j} S_{i,j}S_{i,j} -{1\over L^2}(\tilde x_i)^2+{1\over H}(\tilde x_i\tilde p_i+
\tilde p_i\tilde x_i)-{1\over M^2}(\tilde p_i)^2+({1\over L^2M^2}-{1\over H^2})(f,l)^2
$$
where the generator of the "spin" variable is defined as $S_{i,j}=(I-{ih\over H})F_{i,j}+p_ix_j-
p_jx_i$ and four-dimensional vectors $\tilde x_i,\tilde p_i$ ( of the pseudo co-ordinates and momenta) are defined by $\tilde x_i=\sum \epsilon_{ijkl}x_jF_{kl},\tilde p_i=\sum \epsilon_{ijkl}p_jF_{kl}$. Pseudovectors $\tilde x_i,\tilde p_i$ are introduced in analogy with the Pauli-Lubansky  vector in representation theory of the Poincare algebra. In the classical limit the operators of Casimirs $K_3,K_4$  are fixed quantum numbers of representation of the Lorentz $O(1,3)$ algebra.

By direct calculations using (\ref{2}) it is not difficult to check that $K_3,K_4$  commute with all the generators of the quantum algebra (\ref{3})  

All relations above are written for the algebra of the compact group $O(6)$. But the algebra (\ref{2})
contains as a subalgebra $O(1,3)$ (algebra of Lorence group) and at least one axis in the above formulae must be imaginary. We choose for it the index 4, conserving indices 1,2,3 ($\alpha$) always for  "compact" axes.

The possibility with only the 4-th axis imaginary leads to the $O(1,5)$ algebra with 5 non-compact generators $T_{4,\alpha},T_{6,4},T_{5,4}$. In the case of two imaginary axes, i.e. the $4-th,\ 5-th$ we have the $O(2,4)$ algebra with 8 non-compact generators $T_{4,\alpha},\ T_{5,\alpha},\ T_{6,4}T_{6,5}$. And in the case of 3 imaginary axes, as $4-th,\ 5-th,\ 6-th$ to $ O(3,3)$ algebra with 9 non-compact generators $T_{4,\alpha},T_{5,\alpha},T_{6,\alpha}$.

In conclusion of this introduction we would like to point out that the case of a quantum space with only one dimensional (action) constant $H$ ($L^2\to \infty,M^2\to \infty$) is not contained  among the quantum spaces proposed something about 60 years before in published  papers of Snyder \cite{SN} and Yang \cite{Y}. This case is very intriguing  by the fact that in this case the generators of four co-ordinates are commutative, the group of motion remains the Pouncare one, the theory is relativistic invariant but not  time reversible (and as a consequence $CP$ non invariant). The difference from the usual case consists in thefact that in this case it is not possible to diagonalize four generators of co-ordinates  simultaneously. And thus the field variables can't be represented as a functions of 4 indpendent co-ordinates. In this connection see \cite{FLI}.

For the realisation of such program at first it is necessary to have an explicit representation of the generators of nocompact algebras of six order mentioned above.

\section{Representations of real forms of the algebras of six-dimensional ortogonal groups}

The generators for all nocompact groups $O(p,q)$ was found in \cite{2}. We present below the algebras
applicable for interesting for futher consideration.
 In the general case the algebra of the $O(p,q)$ group consists (in what follows $p\leq q$) of ${(p+q)(p+q-1)\over 2}$ generators.
${p(p-1)\over 2}+{q(q-1)\over 2}$ among them are the compact ones. All other $pq$ generators are non-compact. The maximal compact subgroup coincides with $O(p)\times O(q)$. Matrix elements of the vector representation of orthogonal groups $O(p),O(q)$ are denoted by Greek and Latin indices $p^{\alpha}_{\beta},\quad 1\leq \alpha,\beta \leq p$, $q^i_j \quad 1\leq i,j \leq q$. The generators of regular representations (left and right) of such group are denoted as $P_{\alpha,\beta}, P^{\alpha,\beta}$,$Q_{i,j}, Q^{i,j}$. Generators with upper indices acts only on upper indices of the representation matrix, with the lower ones only on thelower ones
$$
[Q_{i,j},q_k^s]=\delta_{i,k}q_j^s-\delta_{j,k}q_i^s,\quad [Q^{i,j},q_k^s]=\delta_{i,s}q_k^j-\delta_{j,s}q_k^i
$$
And the same follows with respect to Greek indices. Generators of left and right regular representations are connected by equations
$$
Q_{i,j}=\sum_{k,s} q_i^k q_j^s Q^{k,s},\quad Q_{i,j}=\sum_{k,s} q^i_k q^j_s Q_{k,s}
$$
$$
\sum_i q_i^k Q_{i,j}=\sum_s  q_j^s Q^{k,s}
$$
In this notation the compact generators of the "big" $O(p,q)$ algebra coincides with the generators of the left regular representations of the compact algebras. $pq$ nocompact generators appear as
\be
F_{i,\alpha}=\sum_{\rho=1}^p \rho_{\alpha} p^{\mu}_{\alpha} q^{\mu}_i+\sum_{s\leq \nu} p^{\nu}_{\alpha}q^s_i P^{\nu,s}+\sum_{\nu\leq s}p^{\nu}_{\alpha} q^{\mu}_i Q^{\nu,s}\label{1}
\ee
where $\rho^{nu}$ in general are the part of the parameters, determining irreducible representations of the $O(p,q)$ algebra. Using commutation relations between the generators of the compact algebra

$$
[Q_{ij},Q_{sk}]=\delta_{js}Q_{ik}-\delta_{is}Q_{jk}-\delta_{jk}Q_{is}+\delta_{ik}Q_{js})
$$
and the same only with common exchange of the minus sign  in the right side for generators with the upper indices, keeping also in mind that generators of left and right translation  commute it is not difficult (in principle) to check that the proposed route leads to an irreducible representation of the $O(p,q)$ algebra. Except for parameters $\rho_{\alpha}$ irreducible representations of $(O(p,q)$ algebra are defined by $(l_1,l_2,..)$ parameters which define irreducible representations of the compact algebra $O(q-p)$. More detailed information on this subject it is possible to find in the doctoral dissertation of the author \cite{LL} and the published article on the same subject \cite{LS}. 

The Casimir operators for this representation may be calculated explicity with the result
\be
K_2=l_3(l_3+4)+l_2(l_2+2)+l_1^2,\quad K_3=(l_3+2)(l_2+1)l_1,\quad K_4=(l_3+2)K_2+((l_2+1)l_1)^2
\label{KAZ}
\ee
The sense of the introduced parameters $l_i$ will be explained below.

\subsection{The case $O(1,5)$}

In this the Greek indices take only one value and 5 noncomact genrators are enumerated by only one Latin index $1\leq i\leq 5 $
\be 
F_{i,(6)}=\rho q^5_i+\sum_{s=1}^5 q_i^s Q^{(s,5)},\quad F_{i,j}=Q_{i,j}\label{3}
\ee
By the symbol $F$ we denote $15$ generators of the $O(1,5)$ algebra. 5 of them $F_i$ are non-compact
, $10$ $F_{i,j}$ are  compact. Let us pay attention that with respect to the upper index generators $F_i$ have the structure of the $5$-th component of the five dimensional vector
and by thus reason commute with generators of the four dimensional group of rotations connected with $1,2,3,4$ indices. This means that  invariant operators of Casimir which define irreducible representations of $O(1,5)$  the input from the right regular representation of the four-dimensional rotational algebra  may be present only in the form of its two Casimir operators. These two proper values (arbitrary natural or seminatural simultaneously) together with $\rho$ from (\ref{3})  define irreducible representation of $O(1,5)$.   

\subsubsection{Components of the field equations}

The  set of parameters on which corresponding 
representation of six-dimensional algebra is realised  will be called a point in the quantum space. The equation in quantum space arises as aplication of some operators of the algebra to a function depending of the point variables.

The group of motion of $O(1,5)$ quantum space coincides with $O(1,4)$. 10 generators of the algebra of this group are the following ones $p_4,\ f,\vec p,\ l$. The first four are non-compact; 
the last six  are compact. An equation of motion (if we want have the correct "classical" limit)
must be invariant with respect to such transformations (translations and Lorentz transformation in the classical limit).

The main equation describing the motion of the free particle must be invariant with respect to the group of the motion of quantum space generated by infinitesimal displacements and rotations.
Such properties are satisfied by the quadratic d'Alambert operator
$$
\sum_{i=1,2,3,4} p_i^2+\sum_{i\leq j} F_{i,j}F_{i,j}    
$$
And thus the motion of the free particle of the mass $m$ is described in quantum space by the equation

\be
( \sum p_i^2+\sum_{i\leq j} F_{i,j}F_{i,j})\psi=m^2\psi\label{DA}
\ee
But as it was explained above generators $p_i,F_{i,j}$ belongs to a definite representation of the $O(1,5)$ algebra represented above. And thus after substitution of these expressions
(\ref{DA}) we pass to a differential equation of the second order on the ten dimensional compact 
manifold (parameters of group element of $O(5)$).

The additional possibility consists in the introduction of some finite-dimensional (non-compact) representation of $O(1,4)$ algebra with 10 generators $\Gamma$. Then the scalar product of the mutually commutative generators of the  quantum algebra $O(1,4)$  introduced above leads to theequation
$$
(\Gamma_4p_4+(\vec \Gamma_f \vec f)-(\vec \Gamma_p \vec p)-(\vec \Gamma_l \vec l)-m)\psi=0
$$ 
which is obviosly $O(1,4)$ invariant with respect to transformation with infinitesimal 
$p+\Gamma$ generators (total moment).  

Now we are going to present all components of this equation in explicit form with comments afterwards.

In (\ref{DA})  four non-compact generators are $p_4,\vec f$ and six generators $\vec x,\vec l$ are compact. In connection  with the previous subsection for generators of the group of motion 
of $O(1,5)$ quantum space we have
\be
(p_4,\vec f)=\rho n_i^5+\sum n^{\alpha}_i \tilde F^{\alpha,5},\quad (\vec x,\vec l)=F_{ij}
\label{DIF}
\ee 
where latin  indices takes four values $(1,2,3,6)$, the Greek ones five values $(1,2,3,5,6)$;
$n_{\alpha}^{\beta}$ matrix elements of vector representation of $O(5)=B_2$ algebra; $F_{\alpha,\beta},\tilde F^{\alpha,\beta}$ generators of the left (right) regular representation of this algebra connected by $F_{\alpha,\beta}=\sum_{\mu,\nu} 
n_{\alpha}^{\nu}n_{\beta}^{\mu}\tilde F^{\nu,\mu}$.

Now let us parametrize  an element of $O(5)$ compact group by
\be
O(5)=exp T_{12}\tau_1 exp T_{23}\tau_2 exp T_{34}\tau_3 exp T_{45}\tau_4 O(4)\equiv \hat O(5) O(4) \label{PAR}
\ee
and calculate in this parametrization the generators of its left regular representation
$\tilde F^{\alpha,5}$. We have in a consequence
$$
\dot O(5)=\dot {\hat O(5)} O(4)+\hat O(5)\dot O(4)=\hat O(5) O(4)\pmatrix{ 0 & n \cr
                                                                         -n^T & 0 \cr}
$$ 
where $n$ is a four-dimensional vector. After multiplication of the last equation on $ (\hat O(5))^
{-1}$ from the left and on $O(4)^{-1}$ from the right we pass to the matrix system of equations
\be
(\hat O(5))^{-1}\dot {\hat O(5)}+\dot O(4)O^{-1}(4)=\pmatrix{ 0 & (O(4)n) \cr
                                                                 -(O(4)n)^T & 0 \cr}\label{MN}
\ee
Resolving the last system leads to the following expression;
$$
\tilde F^{\alpha,5}=O^{\alpha}_4 \partial_{\tau_4}+O^{\alpha}_3({1\over \sin \tau_4} \partial_{\tau_3}-\cot \tau_4 L_{3,4})+O^{\alpha}_2({1\over \sin \tau_4 \sin \tau_3} \partial_{\tau_2}-{1\over \sin \tau_4}\cot \tau_3 L_{2,3}-\cot \tau_4 L_{2,4})+ 
$$
$$
O^{\alpha}_1({1\over \sin \tau_4 \sin \tau_3 \sin \tau_2} \partial_{\tau_1}-{1\over \sin \tau_4 \sin \tau_3}\cot \tau_2 L_{1,2}-{1\over \sin \tau_4}\cot \tau_3 L_{1,3}-\cot \tau_4 L_{3,4})
$$ 
where $L_{i,j}$ are the generators of the left regular representation. Taking into account 
$\sum_{\alpha} n_i^{\alpha}O(4)^{\alpha}_k=\hat O(5)_i^k$ we obtain for non-compact generators
$\rho n^5_i+\sum n^{\alpha}_i \tilde F^{\alpha,5}=$
$$
\rho \hat O^5_i+\hat O^4_i \partial_{\tau_4}+O^3_i({1\over \sin \tau_4} \partial_{\tau_3}-\cot \tau_4 L_{3,4})+O^2_i({1\over \sin \tau_4 \sin \tau_3} \partial_{\tau_2}-{1\over \sin \tau_4}\cot \tau_3 L_{2,3}-\cot \tau_4 L_{2,4})+ 
$$
$$
\hat O^1_i({1\over \sin \tau_4 \sin \tau_3 \sin \tau_2} \partial_{\tau_1}-{1\over \sin \tau_4 \sin \tau_3}\cot \tau_2 L_{1,2}-{1\over \sin \tau_4}\cot \tau_3 L_{1,3}-\cot \tau_4 L_{3,4})
$$ 
where now $\hat O^{\alpha}_{\beta}$  are matrix elements of matrix $\hat O(5)$ from (\ref{PAR}),
which depend only upon 4 $\tau_i$ co-ordinates. By $L_{i,j}$ are denoted the gnerators of left regular representation of $O(4)$ algebra. But the representation of this algebra is fixed by the condition of irreducibility of the representation of the whole $O(1,5)$ algebra of the quantum space. And thus in the formulae above $L$ are numerical matrices of an arbitrary irreducible $(l_1,l_2)$ representation of the $O(4)$ algebra.

Now it remains only to express compact generators from (\ref{DIF}) in terms of derivatives with respect to  co-ordinates $\tau_i$ and generators $L$.

For this goal it is necessary to modify the right side of the main equation (\ref{MN})
\be
(\hat O(5))^{-1}\dot {\hat O(5)}+\dot O(4)O^{-1}(4)=(\hat O(5))^{-1}\pmatrix{ \omega & 0 \cr
                                                                 0 & 0 \cr}\hat O(5))\label{IMN}
\ee
We present below result of standard not too cumbersome calculations
$$
F_{i,j}=F_{\omega}=\Omega_{3,4} \partial_{\tau_3}+\Omega_{2,4}({1\over \sin \tau_3} \partial_{\tau_2}-\cot \tau_3 L_{2,3})+\Omega_{1,4}({1\over \sin \tau_2 \sin \tau_3} \partial_{\tau_1}-{1\over \sin \tau_3}\cot \tau_2 L_{1,2}-\cot \tau_3 L_{1,3})+
$$ 
$$
\Omega_{2,3}L_{2,3}+\Omega_{1,3}L_{1,3}+\Omega_{1,2}L_{1,2}
$$
where by $\Omega$ is denoted $4\times 4$  $\Omega=(O_4^{-1}\omega O_4),\quad 
O_4=\exp T_{12}\tau_1 \exp T_{23}\tau_2 \exp T_{34}\tau_3$.

Thus from the explicit expressions for generators of irreducible representations of $O(1,5)$ it follows that the equations of motion for free particles are differential equations on a compact manifold
$0\leq \tau_i\leq 2\pi$ or $\pi$. Variables are separable and it is possible to obtain the general solution for free particles of  mass $m^2$ and Lorentz (spin) variables $(l_1,l_2)$.
It is obvious that spectra of the square of the mass will be quantized. The $O(1,4)$ algebra possesses discrete series of unitary representations for  positive values of the second order Casimir operator.

Final comments. The algebra $O(1,5)$ possesses only two series of unitary representations; a
continuous one with $\rho=-2+i\sigma$ and an additional one with    real number $\rho$ such that $-2\leq \rho \leq 0$. Let us once more rewrite the second Casimir operator substituting in it its proper value which was calculated for  the representation considered above in \cite{LMM}
$$
K_2=\nu^2(-4-\sigma^2+l_1(l_1+2)+l_2^2)=({I^2\over \nu^2}+{(x_i-{L^2\over H}p_i)^2\over \nu^2 L^2}+{L^2\over H^2}p_i^2+{(l^2-f^2)\over H^2})
$$
$$
K_3=(-2+i\sigma)(l_1+2)l_1=I(\vec f \vec l)-(\vec f [\vec p,\vec x])-
(\vec l, (p_4\vec x-x_4\vec p))
$$
$$
K_4=(-2+i\sigma)^2(l_1+2)^2+(-2+i\sigma)l_2^2+(l_1+2)^2l_2^2=
$$
$$
\sum_{i\leq j} S_{i,j}S_{i,j} -{1\over L^2}(\tilde x_i)^2+{1\over H}(\tilde x_i\tilde p_i+
\tilde p_i\tilde x_i)-{1\over M^2}(\tilde p_i)^2+({1\over L^2M^2}-{1\over H^2})(f,l)^2
$$
All generators on the left side of the last equality are Hermitian ones. If we want to have a correct classical limit $H^2\to infty$ and this $L^2\to infty,m^2\to infty$ we must have on the right side a positive value. Thus at least the cases $l_1=l_2=0$ are in contradiction with this.
This question about the limiting procedure to classical (non quantised) space need additional consideration.

\subsection{The case of $O(2,4)$}

This case much more rich on unitary representations. Comformal algebra $SU(2,2)=O(2,4)$
posses continues, semidiscrete, discrete and additional serie of unitary representation. All information about this algebra in the technique described above reader can find in \cite{FLII}.
Algebra consists from 8 noncompact generators and 7 compact ones. In this case $\nu^2\leq 0,\quad {H^2\over L^2M^2}\leq 1$ and second Cazimir operator may rewritten in the form 
(we conserve sign plus before the compact generators $\mu^2={H^2\over L^2M^2}-1$)
$$
K_2=\mu^2(-{I^2\over \mu^2}-{(x_i-{L^2\over H}p_i)^2\over \mu^2 L^2}+{L^2\over H^2}p_i^2+{(l^2-f^2)\over H^2})
$$
In the dependence of sign of $L^2$ two possibilities appeares. In the case $0\leq L^2$ the compact are $\tilde x_4=(x_4-{L^2\over H}p_4),\vec p,\vec l$ noncompact ones $I,\tilde {\vec x},
p_4, \vec f$. In this case the group of the motion of the quantum space is $O(1,4)$. In the case
$L^2\leq 0$. The distribution on compact and not compact generators are the following
$(\tilde {\vec x},\vec l,p_4)$ and $(I,\vec p,\tilde x_4,\vec f)$ with the group $O(2,3)$ as the group of the motion. 

Irredusible representation are realised on the space of the parametrs of compact groups $O(4$ and $O(2)$. Matrix elements of vector representation denoted by $q_i^j$. $O(2)$ represented by two dimensional matrix $\pmatrix{ \cos \phi & \sin \phi \cr
                                 -\sin \phi & \cos \phi \cr}$.
     
In this notations explicit expressions for dimensionless components of the observables are the following ones
$$
I=T_{1,4}=\rho^1\cos \phi q^1_4+\rho^2\sin \phi q^2_4+\cos \phi \sum q^i_4 Q^{i,1}+
\sin \phi \sum_{2\leq i} q^i_4 Q^{i,2}+\sin \phi q^1_4 P^{2,1}
$$
$$
\tilde x_4=T_{2,4}=-\rho^1\sin \phi q^1_4+\rho^2\cos \phi q^2_4-\sin \phi \sum q^i_4 Q^{i,1}+
\cos \phi \sum_{2\leq i} q^i_4 Q^{i,2}+\cos \phi q^1_4 P^{2,1}
$$
$$
p_{\alpha}=T_{1,{\alpha}}=\rho^1\cos \phi q^1_{\alpha}+\rho^2\sin \phi q^2_{\alpha}+\cos \phi \sum q^i_{\alpha} Q^{i,1}+\sin \phi \sum_{2\leq i} q^i_{\alpha} Q^{i,2}+\sin \phi q^1_{\alpha} P^{2,1}
$$
$$
f_{\alpha}=T_{2,{\alpha}}=-\rho^1\sin \phi q^1_{\alpha}+\rho^2\cos \phi q^2_{\alpha}-\sin \phi \sum q^i_{\alpha} Q^{i,1}+\cos \phi \sum_{2\leq i} q^i_{\alpha} Q^{i,2}+\cos \phi q^1_{\alpha} P^{2,1}
$$
$$
l_{\alpha}=\epsilon _{\alpha,\beta,\gamma}Q_{\beta,\gamma},\quad p_4=T_{4,2}=P_{2,1}
\quad \tilde x_{\alpha}=T_{\alpha,4}=Q_{\alpha,4}
$$
As reader can see from explicit expressions for all above generators, they contain only upper indexes $1,2$ connected with the four-dimensional group of rotation $\sum q^i_{\alpha} Q^{i,1(2)}$. This means that all generators of $O(2,4)$ algebra are invariant with respect to two dimensional rotation in $3-4$ plane generating by $Q^{3,4}$. Proper value of this generator $l_3$ together with $\rho^1,\rho^2$ determines the irreducible representation of $O(2,4)$ algebra. 

Kazimir operators for this representation are the following ones
$$
K_2=\rho^1(\rho^1+4)+\rho^2(\rho^2+2)-l_3^2,\quad K_3=i(\rho^1+2)(\rho^2+1)l_3,\quad 
$$
\be
K_4=(\rho^1+2)^2((\rho^2+2)^2-l_3^2)-(\rho^2+1)l_3)^2\label{KAZI}
\ee
Thus irreducible representation $(\rho^1,\rho^2,l_3)$ of $O(2,4)$ algebra 
is realized on six compact parameters and all equations in such versia of quantum space will be partial differential equations in six-dimensional space.

\subsection{Equation of free motion. $O(2,3)$ case}

Equation describing the motion of the free particle is the ussual one (it present below in dimesionless notations)
$$
((\vec f^)2+(\vec p)^2-p_4^2-(\vec l)^2)\psi+m^2\psi=0
$$

Result of the substitution in it of all corresponding elements from formulae above leads to
$$
[\rho_1(\rho_1+4)(1-(q^1_4)^2)+\rho_2(\rho_2+2)(1-(q^2_4)^2)-(\rho_1+4)q^1_4 M+\rho_1 M^T q^1_4-(\rho_2+2)q^2_4N+
$$
$$
\rho_2 N^Tq^2_4+q^1_4(N-N^T)it+2(\rho_2+1)q^1_4q^2_4it+(q^1_4)^2t^2+N^TN+M^TM+\sum  Q^{i,1}Q^{i,1}+
$$
$$
\sum_{2\leq i} Q^{i,2}Q^{i,2}-(Q_{23}Q_{23}+Q_{21}Q_{21}+Q_{31}Q_{31})
$$
where $\frac{\partial \psi}{\partial \phi}=it\psi$ and $M=\sum q^i_4 Q^{i,1},N=\sum q^i_4 Q^{i,2}$.

\subsection{Equation of free motion. $O(1,4)$ case}

In this case 6 generators $(\vec p,\vec l)$ of $O(1,4)$ algebra are compact ones and $(p_4,\vec f)$ are noncompact. This means that in the general formulas of representation it is only necessary to change by the places $\tilde x_i\to p_i$. Using this prescription we obtain equation of the free motion in the form
$$
(\tilde x_4^2+(\vec f)^2-(\tilde {\vec x})^2-(\vec l)^2)\psi=m^2\psi
$$
By direct substitution of above expressions and some not combersom calculations for the operator of mass $m^2$ we obtain
$$
\rho_1(\rho_1+4)\cos^2 \phi+\rho_2(\rho_2+2)\sin^2 \phi +\cos^2 \phi(Q^{21})^2+\sum_{3\leq j}
(\cos \phi Q^{j1}+\sin \phi Q^{j2})^2-2(\rho_2+2)\sin \phi \cos \phi Q^{21}-
$$
$$
(\rho_1+4)\sin \phi \cos \phi \partial_{\phi}+2(\rho_2+2)\sin \phi \cos \phi Q^{21}+
\partial_{\phi}\sin^2 \phi \partial_{\phi}-\sum_{i\leq j}Q^{ij}Q^{ij}
$$
After canonical transformation $m^2\to \exp -\phi Q^{21} m^2 \exp \phi Q^{21}$ the last operator looks as
$$
\rho_1(\rho_1+4)\cos^2 \phi+\rho_2(\rho_2+2)\sin^2 \phi+\partial_{\phi}\sin^2 \phi \partial_{\phi}+\partial_{\phi}\sin^2 \phi
 Q^{21}+\sin^2 Q^{21}\phi \partial_{\phi}+
$$
$$
2(\rho_1+\rho_2+3)\sin \phi \cos \phi Q^{21}-2(\rho_2+2)\sin \phi \cos \phi Q^{21}-(\rho_1+4)\sin \phi \cos \phi \partial_{\phi}-(Q^{23}Q^{23}+Q^{24}Q^{24}+Q^{34}Q^{34})
$$

\subsection {The case of $O(3,3)$}

Continuous series of unitary representations of this algebra  are defined by three complex numbers
$l_3=\rho_3=-2+i\sigma_3,\ l_2=\rho_2=-1+i\sigma_2,\ l_1=\rho_1=i\sigma_1$.

The second Casimir operator for this representation is the following (see (\ref{KAZ}))
$$
K_2=4+\sigma_3^2+1+\sigma_2^2+\sigma_1^2
$$ 
But in this case the identity operator $I$ is  compact  and its square arises on the right hand side of $K_2$ with  negative sign, This means that in this case passing to the limit of the usual theory is impossible using the unitary representation of the $O(3,3)$ algebra. For this reason for consideration of this case requires some additional ideas, yet to be discovered.

\subsection{Degenerate case ${H^2\over L^2M^2}=1$}

In this case the second Casimir operator takes the form
$$
K_2=I^2+{(x_i-{L^2\over H}p_i)^2\over L^2}
$$
and all generators in it are mutually commtative. 
$$
[x_i-{L^2\over H}p_i,x_j-{L^2\over H}p_j]=ih({1\over M^2}-{l^2\over H^2})F_{i,j}=0
$$
In the same way it is not difficult verify that
$$
[I,x_i-{L^2\over H}p_i]=ih({1\over M^2}-{l^2\over H^2})p_i=0
$$
This corresponds to the case of semisimple algebras $O(1,4),O(2,3)$ algebras with translations. The irreducible unitary representations of such algebras may be constructed in complete analogy with the well known case of Poicare algebra. 
Five proper values of mutually commuting generators from the Casimir operator above may be considered as a basis of a representation. Then the representation is realized on  functions of four independent co-ordinates (for definitnes $0\leq L^2$)
$$
I=\cosh \sigma \cos \theta,\quad \vec x-{L^2\over H}\vec p_i=\cosh \sigma \sin \theta \vec n, 
,\quad x_4-{L^2\over H}p_4=\sinh \sigma
$$
where $(\vec n)^2=1$. Generators of the group of motion $O(1,5)$ coinside with the usual genrators of this algebra in a realization on one hyperboloid. Explicit expressions may obtained from after exchange of the corresponding angle with an imaginary one in the formulae for the $O(5)$ algebra presented above (see subsection 2).

And the last comment. In the case of the space of constant curvature the situation is the similar one. The group of motion coinside with de-Sitter groups $O(1,4),O(2,3)$. And this case selfconsistence theory exists in the framework of the ussual theory. Situation with Snyder quantum space is exactly the same as in this subsection.

\section{Gauge theory in quantum space}

The main idea of all gauge theories consists in the construction of some object $F$ depending on the point of space-time, which under gauge transformation $G$ transforms as $F\to G^{-1} F G$. In the case of the quantum space considered above the group of the space motion coincides with $O(1,4)$ or $O(2,3)$. Thus gauge potential must have the same transformation properties and the tensor of the gauge field may constructed in following manner
$$
F_{ij,kl}=[T_{ij}-A_{i,j},T_{k,l}-A_{k,l}]-
$$
\be
g_{j,k}(T_{i,l}-A_{i,l})-g_{i,k}(T_{j,l}-A_{j,l})-g_{j,l}(T_{i,k}-A_{i,k})+g_{i,l}(T_{j,k}-A_{j,k})\label{TEN}
\ee
where $T_{i,j}$ are the generators of $O(5)$ algebra of the quantum space and $A_{i,j}$
the gauge potential taking values in some compact algebra.
From the explicit form of the right side of (\ref{TEN}) it follows immediately that $F_{ij,kl}$
is a function of the point of the quantum space (all generators $T_{m,n}$ arising  are mutualy cancelled) and under the gauge transformation tensor of the field $F_{ij,kl}$ has correct transformation properties. The number of independent components of field tensor is equal to $100\times N_G$ ($N_G$ the dimension of the compact gauge algebra).  It is natural to connect four gauge potentials $A_{i,5}$ with electromagnetic potentials, six other ones $F_{i,j}$
with gauge potentials of the gravitation field. We would like to point out here  that long ago  attention was payed to the $O(5)$ symmetry of electromagnetism and gravity. But the question always arose as to what to do with the additional fifth co-ordinate \cite{KK}, \cite{RUM}. 
In the case of the quantum space this difficulty is absent. Point in quantum space is defined as
the number of continues parameters on which representation of real form of $O(6)$ is realized.
In the cases considered above, the number of continuous parameters is four in the case of the $O(1,5)$ quantum space, maximal six in the $O(2,4)$ one and lastly four in the degenerate case of the previous section. Thus on the background of (\ref{TEN}) it is possible to construct a unitary theory of  gravitation and magnetism in quantum space. Of course this will not be a metrical theory of Einstein type ( in this connection see \cite{LM}). But without detailed consideration it is not profitable to speculate on this subject.

\section{Outlook}

We have not made any attempt to give for all formulae above any physical interpretation.
Of course this is not a simple problem but it was out of the scope of the present paper.
Our goal was to show that it is possible to construct a consistent mathematical formalism to
include into consideration quantum spaces. What we have in a result? We present explicit expressions for all ingredients nesessary for the construction of a consistent field formalism. We have an explicit form for operators of the group of motion of the quantum space necessary for the constuction of the field theory. As in the usual case particles  interract via a gauge field
equation for which  follows directky from (\ref{TEN}). The equations for the particles are invariant with respect to $O(1,4)$ or $O(2,3)$ group of motion. This demand uniquely fixes the form of equations. We have preserved all symmetry properties of the ussual theory but have no answer on the question what is the nature of the space-time. Really situation in micro and macro distances is the same in the today theory. On experiments we observe directions and energy-impulses charecteristics in the point of observation and after this do some deductive conclusions what have happened and in what point of the real space-time. Thus formalism of the present paper must be added by some phylosofical sheam of the process of the mesurement.        

\section{Acknowlegements}

Author is indepeted to D.B.Fairlie for fruithfull disscusion in the process of the 
writing of the present paper and big help in preparation the manuscript for publication.

Author is indepted to CONNECIT for partial financical support.

\section{Appendix I}

In this appendix the case of $(1+1)$ quantized subspace is considered in detail. The algebra of the quantum space coincides with $O(1,3)$ or $O(2,2)$ in  the dependence of signs of dimensional constants.

\subsection{The $O(2,2)$ case}. 

The condition on dimensional constants ${H^2\over L^2M^2}\leq 1$ where the signs of $M^2,\ L^2$ are arbitrary, corresponds to the general $O(2,4)$ case. 
In connection with the general results described in section 2 we have the following expressions for generators ( for definitness $0\leq M^2,\ 0\leq L^2$).
$$
f=h(\cos \phi_2(\rho_1\cos \phi_1-\sin \phi_1\partial_{\phi_1})+\sin \phi_2(\rho_2\sin \phi_1-\cos \phi_1\partial_{\phi_2}))
$$
$$
p={h\over L}\partial_{\phi_1},\quad p_4={h\over L}(\cos \phi_2(\rho_1\sin \phi_1+\cos \phi_1\partial_{\phi_1})-\sin \phi_2(\rho_2\cos \phi_1\sin \phi_1\partial_{\phi_2}))
$$
$$
x={h \nu L\over H}(\sin \phi_2(\rho_1\cos \phi_1-\sin \phi_1\partial_{\phi_1})-\cos \phi_2(\rho_2\sin \phi_1-\cos \phi_1\partial_{\phi_2}))+{h L\over H}\partial_{\phi_1}
$$
$$
x_4-{L^2\over H}p_4={h \nu L\over H}\partial_{\phi_2}
$$
$$
I={h \nu\over H} (-\sin \phi_2(\rho_1\sin \phi_1+\cos \phi_1\partial_{\phi_1})-\cos \phi_2(\rho_2\cos \phi_1\sin \phi_1\partial_{\phi_2}))
$$

\section{Appendix II}

 Below are presented three-dimensional components from which the Casimir operators from section 2 are constructed
$$
S_{\alpha,4}=s^1_{\alpha}, S_{\alpha,\beta}=\sum_{\gamma} \epsilon_{\alpha,\beta,\gamma} s^2_{\gamma}
$$
$$
\vec s^1=(I-i{h\over H})\vec f+\vec p x+4-p_4\vec x,\quad \vec s^2=(I-i{h\over H})\vec l+\vec p x_4-p_4\vec x, \quad \tilde x_4=(\vec x\vec l),\quad \tilde p_4=(\vec p\vec l),
$$
$$
\tilde {\vec x}=x_4 \vec l+[\vec x,\vec f],\quad \tilde {\vec p}=p_4 \vec l+[\vec p,\vec f]
$$
$$
(\rho \sin \nu+({1\over \delta} -\cos \nu)p_{\nu})^2+({1\over \delta} -\cos \nu)^2{L^2\over \sin^2 \nu}=E=
$$

\end{document}